\documentclass[titlepage, 12pt]{article}
\usepackage[margin=1in]{geometry} 
\usepackage{amssymb} 
\usepackage{amsmath} 
\usepackage{amsthm} 
\usepackage{graphicx} 
\usepackage{booktabs} 
\usepackage{natbib} 
\usepackage{pgfplots} 
\usepgfplotslibrary{fillbetween} 
\usepackage{bm} 
\usepackage{xcolor} 
\definecolor{royalblue}{HTML}{0000CD} 
\usepackage[colorlinks=true]{hyperref} 
\hypersetup{
    colorlinks=true,
    allcolors=royalblue,
} 
\usepackage{threeparttable} 
\usepackage{thmtools} 
\allowdisplaybreaks 
\usepackage{mathtools} 

\theoremstyle{definition}

\providecommand{\keywords}[1]
{
  \noindent\small	
  \textbf{\textit{Keywords:}} #1
}

\usepackage{caption} 
\usepackage{subcaption} 

\usepackage{setspace} 
\usepackage{tabularx}
\usepackage{rotating} 

\usepackage{adjustbox}
\usepackage{array}

\newcolumntype{R}[2]{%
    >{\adjustbox{angle=#1,lap=\width-(#2)}\bgroup}%
    l%
    <{\egroup}%
}
\newcommand*\rot{\multicolumn{1}{R{45}{1em}}}

\usepackage{url}

\begin{document}

\title{A New Framework to Estimate Return on Investment for
Player Salaries in the National Basketball Association}

\author{
  Jackson P. Lautier\footnote{Department of Mathematical Sciences,
  Bentley University}\thanks{Corresponding to jlautier@bentley.edu.}
}

\date{\today}

\maketitle

\begin{abstract}
The National Basketball Association (NBA) imposes a player salary cap.  It is
therefore useful to develop tools to measure the relative realized return of a
player’s salary given their on court performance.  Very few such studies exist,
however. We thus present the first known framework to estimate a return on
investment (ROI) for NBA player contracts.
The framework operates in five parts.  First, we decide on a measurement time
horizon, such as the standard 82-game NBA regular season.  Second, we propose
the novel  game contribution percentage (GCP) measure, which is a single game
summary statistic that sums to unity for each competing team and is comprised
of traditional, playtype, hustle, box outs, defensive, tracking, and rebounding
per game NBA statistics.  Next, we estimate the single game value (SGV) of each
regular season NBA game using a standard currency conversion calculation.
Fourth, we multiply the SGV by the vector of realized GCPs to obtain a series
of realized per-player single season cash flows.  Finally, we use the player
salary as an initial investment to perform the traditional ROI calculation.
We illustrate our framework by compiling a novel, sharable
dataset of per game GCP statistics and salaries for the 2022-2023 NBA regular
season.  Using only total GCP, we find the top five performers to be Domantas
Sabonis, Nikola Joki\'{c}, Joel Embiid, Luka Don\v{c}i\'{c}, and Bam Adebayo.
With player
salaries, however, the top five ROI performers become Tre Jones, Kevon Harris,
Nick Richards, Ayo Dosunmu, and Max Strus.  A scatter plot of ROI by salary for
all players is presented.
Notably, missed games are treated as defaults because GCP is a per game metric.
This allows for break-even calculations between high-performing players with
frequent missed games and average performers with few missed games, which we
demonstrate with a comparison of the 2023 NBA regular seasons of Anthony Davis
and Brook Lopez.
We conclude by suggesting uses of our framework,
discussing its flexibility through customization, and outlining potential
future improvements.

\bigskip


\end{abstract}

\doublespacing

\section{Introduction}

On December 20, 2022, the Phoenix Suns of the National Basketball Association
(NBA) in combination with the Phoenix Mercury of the Women's National
Basketball Association were purchased at a valuation price of \$4 billion
\citep{woj_suns}; the NBA is big business.  As in any financial operation, it
is of great interest to assess performance for the purposes of allocating
returns to investors or other parties with pecuniary interests. Among the
many financial interests of NBA teams, such as ticket sales, television
revenue, and merchandise sales, there is the obvious consideration of
compensating the players that make up each team's roster.  This task is made
complicated by a myriad of reasons, not the least of which is that the NBA
operates within a framework designed to restrict a free market.  Indeed, a
crucial component of the competitive parameters of the NBA is a {\it salary
cap}, which set a minimum total salary of \$111.290 million and a maximum
total salary of \$123.655 million per team for the 2022-2023 NBA season
\citep{nba_salary_cap}, subject to numerous additional restrictions
\citep{nba_cba}.  Thus, how to effectively allocate this fixed total salary to
on court personnel is a crucial component of a team's on court success.

It is natural, then, to suppose there exists a great number of studies
that consider both on court performance and salary simultaneously to arrive
at methods to measure realized return on investment (ROI) or the internal
rate of return (IRR) of a player's contract in view of said player's on
court performance.
A survey of related studies indicates that this is not the case, however.
\citet{idson_2000} attempt to derive the determinants of a player's salary
in the National Hockey League with a model that incorporates the performance of
teammates.  We consider the NBA, however, and our methodology differs
considerably (see Section~\ref{sec:methods}).
\citet{berri_2005} identify the importance of height in the NBA
and juxtaposes it against population height
distributions to explain competitive imbalances observed in the NBA.  Such
imbalances are thought to negatively impact economic outcomes of sports leagues
\citep{berri_2005}.  While financial considerations enter into the analysis
of \citet{berri_2005}, it does not concern the ROI of single players but rather
professional leagues overall.
\citet{tunaru_2005} develop a claim contingent framework that is connected to
an option style valuation of an on field performance index for football (i.e.,
soccer) players.  Our proposed method differs materially, however, and we
focus on basketball rather than football.
\citet{berri_2006} find mixed results to the question of whether or not
signing a long-term contract leads to shirking behavior from NBA players.  The
overall objective of their study differs meaningfully from that of our proposed
realized ROI metric, however. More recently,
\citet{simmons_2011} find salary inequality is effectively independent of
player and team performance in the NBA, a result that runs counter to the
hypothesis of fairness in traditional labor economics literature.
In a related study, \citet{halevy_2012} find the hierarchical structure of pay
in the NBA can enhance performance.  Neither study attempts to produce a
contractual ROI, however.
\citet{kuehn_2017} assumes the ultimate goal of each team is to maximize their
expected number of wins to find teammates have a significant impact on an
individual player's productivity.  \citet{kuehn_2017} subsequently reports
that player salaries are determined instead mainly by individual offensive
production, which can lead to a misalignment of incentives between individual
players and team objectives.  Of note, the salary findings of
\citet{kuehn_2017} correspond to those of \citet{berri_2007}, a similar
study.  Our forthcoming analysis differs from all of these studies
generally in that we do not attempt to explain salary decisions and
instead propose the first known framework to measure the realized return of
a player's contract in light of on court performance.

We do so by translating a player's series of recorded regular season games
in terms of on court production (i.e., box score statistics, player tracking
data, rebounding statistics, etc.; see Table~\ref{tab:fields}) into a
series of realized cash flows.
This is done by our novel game contribution percentage
(GCP) player evaluation metric and a standard currency conversion calculation.
By treating the player's salary as the initial time zero investment and using
the converted series of regular season games into realized cash flows, we are
able to calculate the realized return using traditional financial tools
\citep[e.g.,][]{berk_2007}. We
note that the framework we propose considers all single game units separately,
the importance of which is well-known in canonical basketball treatments
\citep[e.g.,][Chapter 16, pg. 192]{oliver_2004}.  In other words, our approach
treats any missed game as a zero cash flow, which implicitly allows us to
quantify the cost of missing games (e.g., Figure~\ref{fig:gcp_comp}).
Furthermore, our GCP metric is directly
comparable between players, and it does not require standardization because it
is a per game metric (and thus already standardized).  Finally, while we
propose a contractual ROI framework that is meant to be used directly, we also
carefully note the ways the method may be altered or enhanced to meet the needs
of future analysts (e.g., the investment time horizon need not be one regular
season).  That is, it is the framework we propose that is our main
contribution, and we take care to offer suggestions for customization.

We proceed as follows.  The bulk of the effort is Section~\ref{sec:methods}, in
which we first define and illustrate the GCP method in
Section~\ref{subsec:GCP_method}, as well as contextualize its novelty within
the current landscape of basketball statistics.
Next, we demonstrate how GCP may be used to
complete a financial realized ROI calculation for each player in
Section~\ref{subsec:ROI_method}.
Section~\ref{sec:results} then performs the realized calculations we propose
for all players from the 2022-2023 NBA regular season.  We first discuss the
creation of a novel dataset that combines NBA player tracking data
\citep{nba_stats} with player salary information \citep{hoops_hype}.
We then present cumulative GCP results for the 2022-2023 NBA regular season,
irrespective of player salary, in comparison with other popular advanced NBA
metrics to demonstrate the utility of the novel GCP outside of the ROI
framework.  Section~\ref{subsec:ROI_results}
then performs the realized ROI calculations, and we report the top and bottom
50 performers, including a scatter plot of ROI by salary.
In support of reproducible research,
the complete compiled data and replication code may be found on a public
\texttt{github} repository at
\href{https://github.com/jackson-lautier/nba_roi}{https://github.com/jackson-lautier/nba\_roi}.  Finally,
the manuscript concludes with Section~\ref{sec:disc},
which discusses how our methodology may be utilized by
NBA player personnel decision makers, NBA award voters, and NBA governing
bodies before closing with comments regarding a number of possibilities to
customize and improve upon our proposed ROI framework.

\section{Methods}
\label{sec:methods}

This is a lengthy two-part section.  In Section~\ref{subsec:GCP_method}, we
introduce the GCP methodology.  We justify why a new metric is necessary
despite a crowded landscape of on court statistical evaluation methods,
provide a formal definition and justification of our approach,
and close with an illustrative calculation.
In Section~\ref{subsec:ROI_method}, we build on the work of
Section~\ref{subsec:GCP_method} with player salary data and standard currency
conversation economic calculations to demonstrate how the realized ROI may
then be calculated using a financial cash flow framework.

\subsection{Game Contribution Percentage}
\label{subsec:GCP_method}

Recall the overall objective of our framework: converting a player's
series of recorded games in terms of on court performance into a series of
realized cash flows.
Because we may estimate the dollar value of a single NBA game (see
Section~\ref{subsubsec:currency}), i.e., a single game value (SGV), it is left
to allocate the SGV to each of the game's active
players.  That is, a theoretically ideal measure reports the true percentage
contribution of each player per game (i.e., a
{\it game contribution percentage}).
If we had such a measure, we could then imagine the
counterfactual of each team multiplying the SGV by the percentage contribution
to find the fair amount of financial compensation earned by each of the game's
participants, considering only on court performance.
Repeating this calculation
for all recorded games over the chosen investment horizon will therefore yield
a series of CFs for all players, as desired.
For ease of exposition, we will assume the investment horizon to
be the standard NBA regular season (i.e., we desire to produce a series of 82
cash flows for each player).  This is for illustration only; the measurement
time horizon is a flexible input into the framework.

Hence, we restrict all calculations to a contained single game unit.
Indeed, the importance of the single game unit is well-known
\citep[e.g.,][Chapter 16, pg. 192]{oliver_2004}, and it
is thus the most natural delineation of NBA performance units.
Further, each game is treated as a separate, contained entity by NBA league
rules.  In other words, margin of victory has no
bearing on regular season standings in the NBA as of this writing.
This is a nuanced point that warrants emphasis.  We are interpreting a player
contract as a debt instrument of a single investment (i.e., the player salary)
that obligates the player to produce 82 per game payments.  Because a player
cannot contribute any more than the entire SGV within a single game,
missed games are
treated as {\it defaults} or missed payments in our framework.
(This means that running season totals of GCP, such as those discussed in
Section~\ref{subsec:PVGCP_results}, allow analysts to determine the exact
inflection point of a dominant player that misses many games versus a solid
player that consistently plays; e.g., Figure~\ref{fig:gcp_comp}.)
In a further minor point,
limiting calculations to a single game and working in terms of
percentages also helps somewhat offset issues from {\it garbage time}
\citep[e.g.,][pg. 138]{oliver_2004}, which may result in players inflating
their statistics, or the need for {\it per possession} standardization
\citep[e.g.,][pg. 25]{oliver_2004}.

Is a new metric necessary?  Given what is available at present, we believe
the answer is affirmative. Classical regression treatments, such as
\citet{berri_1999}, do not perform calculations on a game-by-game basis and
have become dated in light of the advancements in data availability
\citep{nba_stats}.  Data advancements also rule out \citet{page_2007},
who fit a hierarchical Bayesian model to 1996-1997 NBA box
score data to measure the relative importance of a position to winning
basketball games.  The same is true for \citet{fearnhead_2011}, who,
in another Bayesian study, propose an NBA player ability assessment model
that is calibrated to the relative strength of opponents
on the court (via various forms of prior season data; \citet{fearnhead_2011}
provide results for the 2008-2009 NBA regular season).
The work of \citet{casals_2013}, who fit an ordinary least squares (OLS)
model to 2006-2007 NBA
regular season data in an attempt to measure the game-to-game variability of
a player's contribution to points and win score \citep{berri_2010}, is
closer in spirit but does not provide the level of box score detail we
require.
In a promising study, \citet{lackritz_2021} create a model to assign fractional
credit to scoring statistics for players in the NBA.  Unfortunately,
\citet{lackritz_2021} consider only offensive statistics.
Finally, the aforementioned \citet{idson_2000} and \citet{tunaru_2005} do not
consider basketball.

It also worth considering popular basketball metrics, such as
{\it NBA Win Shares} \citep{win_share} based on \citet{oliver_2004} or those
summarized in Table~\ref{tab:top50_GCP} from \citet{nba_leaders}.  Despite
the fair criticism of \citet{berri_2010}, these metrics deserve consideration
given their popularity.  In particular, {\it Game Score} \citep{br_glossary}
appears to exactly meet our needs.  Upon review, however, Game Score does not
utilize any of the NBA data advancements of Table~\ref{tab:fields} and also
utilizes hard-coded coefficients, which are difficult to interpret generally.
Further, the popular metrics included in Table~\ref{tab:top50_GCP} are
calculated as a running total of cumulative statistics and thus do not
adequately meet our game-by-game needs.  Hence, we believe adding GCP to the
growing pile of basketball statistics is justified.

There is an admitted level of subjectivity to assigning credit to players in a
basketball game.  \citet[][Chapter 13]{oliver_2004} provides a nice
introduction to this problem, though our approach differs materially from his
{\it Difficulty Theory}.  Indeed, we make no attempt to claim that the GCP
metric we propose is perfect and would even go so far as to concede we have
intentionally erred on the side of simplicity so as not lose sight of the
overall ROI framework design.
Nonetheless, we do anticipate a close read of
our reasoning behind the principles of the GCP will alleviate concerns it is
too artificial.  To reiterate: the purpose of GCP is to
illustrate the percentage credit calculation necessary to perform our
realized contractual ROI calculation for NBA players.  Thus, the GCP method may
be tweaked, updated, or overhauled by future analysts
without materially changing
the framework herein, as long as any future contribution percentage measure
sums to unity.  In short, customization is possible, and we will elaborate on
this point in Section~\ref{sec:disc}.
As a final comment before proceeding to formally introduce GCP,
\citet{terner_2021} provide a comprehensive review of the current landscape of
basketball statistics.  Hence, one may review \citet{terner_2021} to
see that the GCP concept, especially for the combination of statistics we
propose in Table~\ref{tab:fields}, is itself a contribution to the basketball
analysis literature.

\subsubsection{Definition}

In attempting to measure a nebulous theoretical construct, such as a basketball
player's GCP, it is unavoidable to first establish a set of fundamental
principles or axioms.  In light of the objective of
this manuscript, which is to establish a working framework to estimate a
player's realized ROI and simultaneously provide a benchmark calculation,
we make a good faith attempt at instituting the following six principles:
value all activity, process over results, no double counting,
venerate the {\it fifty-fifty ball}, sign and affect agnostic, and
retrospective over prospective.  We now discuss each in turn.

{\it Value All Activity}.
We desire to recognize any form of on court activity. This is in
deference to the truism that it is possible to impact a basketball game without
recording traditional box score statistics.  A classical example is a defensive
player that contests a shot to the point it results in an altered, inaccurate
field goal attempt, but the defender does not tip the ball.  In this case, the
defender's contest had an impact but a traditional {\it block} would not be
recorded. Therefore, in addition to the
traditional statistical categories, such as {\it two-point field goals made},
{\it turnovers}, and {\it blocks}, we also utilize more recent player
tracking and hustle statistics, such as {\it distance traveled},
{\it box outs},
and {\it touches}.  This principle is also why we calculate a GCP for both
the winning and losing team.  Quite simply, the zero-sum nature of wins and
losses in an NBA regular season suggests that approximately 50\%
of all player compensation is for losses.
Thus, we desire to recognize losses in our novel ROI framework.  Notably,
this differs significantly from a wins-focused analysis
\citep[e.g.,][]{win_share} and also eliminates
the Factors Determining Production of \citet{martinez_2012}, which is a model
based on non-scoring box score statistics and is fitted via OLS against the
difference in final score.

{\it Process Over Results}.
We desire to recognize the virtue of a single player's individual
process over the resulting outcome.
This is admittedly a controversial position, even without wading into the now
infamous period in the history of the \texttt{Philadelphia 76ers}
\citep{the_process}.  Our
reasoning stems from a preference for virtue-based (or character) ethics
over outcome-based or duty- and rule-based ethics.  In other words, under
this principle, a good
outcome, like a made basket, does not absolve a potentially poor decision, like
shooting against a triple-team and ignoring open teammates. For an excellent
introduction to such ideas within the context of economics, see
\citet{wight_2015}.  Within a basketball game, a classical example would
be a player that makes an excellent pass to a teammate for a high-percentage
look at the basket (good process), but the recipient of the pass misses the
shot (bad outcome).  In this example, the traditional {\it assist} statistic
would not be recorded because there was not a made basket. In addition, the
passing player, aside from delivering a quality pass, has no control on the
receiving player's ability to make the basket. Hence, we prefer the
statistic {\it potential assists} to the traditional assists.  Similarly,
we prefer an adjusted form of
{\it rebound chances} to the traditional {\it rebounds},
and we track both field goals made and field goals missed. In some instances,
we are unfortunately constrained by data availability.  For example, it is
preferable to track screens set instead of {\it screen assists}, but detailed
data for screens set by game is not readily available as of this writing.

{\it No Double Counting}.
We desire to avoid the classical economics problem of {\it double counting},
which is undesirable in the measurement of macroeconomic calculations
like {\it gross domestic product} \citep[e.g.,][Chapter 10]{mankiw_2003}.
In essence, our objective is to avoid giving a player double credit in the
GCP calculation.  In some cases, the adjustments are
straightforward.  For example, we create statistics such as three-point field
goals missed rather than use both three-point field goals made and three-point
field goal attempts, and we track potential assists but do not also include the
traditional assists (see the discussion in the principle {\it Process Over
Results} to see why we do not differentiate between assists and potential
assists).  Similarly, we track two-point field goals made,
three-point field goals made, and free throws made but do not also track total
points scored.  In other instances, we make some subjective adjustments.
For example, we subtract {\it contested rebounds} from {\it rebound chances},
and we subtract blocks from {\it contested two-point shots}.  For the latter,
it is possible a three-point shot was blocked, but we make the assumption most
blocked field goal attempts are two-point shots.  Lastly, we subtract both
potential assists and {\it secondary assists} from {\it passes made}. In
other instances, it is more difficult to parse out possible double counting.
For example, we track both {\it minutes played} and {\it possessions played}.
Conceptually, these categories track similar metrics and must overlap or
double count in some form.  An adjustment is not straightforward, however, and
so we elect to track both at present.

{\it Venerate the Fifty-fifty Ball}.
Given the importance of each possession in a basketball game, we make an effort
to recognize moments when possession of the ball is uncertain.  This is clear
in our use of {\it loose balls recovered} within the GCP calculation.  More
subtle perhaps, is our preference to track contested rebounds over rebounds.
In the moment a field attempt is missed, future possession is uncertain. Hence,
we find it is of more value to record a rebound when it is contested then when
the offensive team does not elect to pursue the ball.  Indeed, the value of
such rebounds and possession in general is well understood within the context
of valuing a basketball player's contribution
\citep[e.g.,][Chapter 2, 6]{oliver_2004}, and so we omit additional
explanation. We further acknowledge our preference to differentiate between
rebounds and contested rebounds borrowers from the aforementioned
{\it Difficulty Theory} \citep[][Chapter 13]{oliver_2004}.

{\it Sign and Affect Agnostic}.
This principle perhaps differs the most from traditional basketball player
evaluation metrics, such as {\it win shares} \citep{oliver_2004, win_share}
or {\it game score} \citep{br_glossary}.
In short, we do not distinguish between a positive and
negative contribution, and we do not attempt to measure the relative value or
impact of one statistical category versus another.  While such a principle does
contribute towards a final metric that is easy to interpret and easily tweaked
to meet the specifications of different analysts, i.e., suitable to establish
a framework, we do also feel justification
is possible.  Consider first the traditional statistic turnovers.  It is nearly
unanimous to basketball analysts that loosing possession of the basketball is a
negative outcome.  Our GCP metric does not attempt to measure a player's
contribution to winning, however.  Instead, it is more akin to
\texttt{usage percentage} \citep{nba_glossary} in that we attempt to measure a
player's overall contribution to a game's outcome.  From this point-of-view, it
is not unreasonable to suggest that a player with many turnovers in a single
game likely had a large contribution on the outcome, albeit negative. In terms
of relative impact, difficulties quickly arise in attempting to assign relative
value.  For example, it appears straightforward that a made three-point field
goal should be worth 50\% more than a made two-point field goal.  Basketball is
more nuanced, however.  If a player has the ability to easily score two-point
field goals near the basket, then the defense must adjust their approach with
double teams.  This in turn will leave other players open, which may lead to
valuable open field goal attempts.  We thus elect to use an equal weighting
system as both a logical starting point and for ease of interpretation.
This principle rules out many current basketball player contribution statistics
already discussed.  One not yet mentioned and also not suitable for our needs,
however, is
\citet{niemi_2010}, who offers a hierarchical model to derive underlying
distributions for player contributions and considers play-by-play data
from the 2009-2010 NBA regular season.

{\it Retrospective Over Prospective}.
Finally, we remain aligned with the principles of financial accounting in that
we consider only what was actually received on the ledger.  In other words, we
do not adjust for potential randomness.  For statistically minded readers, this
may appear troubling.  Indeed, for the purposes of designing an offense, for
example, it is more valuable to know the long-term average field goal
percentage of a shot location than if a player happened to make or miss one
single field goal attempt.  This differs from our objective, however, and we
illustrate with an example from consumer finance.  If a borrower misses a
monthly mortgage payment, it does little for the lender to hear an explanation
that similar borrowers made last month's payment with a high percentage on
average.  From the lender's perspective, it only matters that the payment was
missed.  Hence, as a form of retrospective accounting, we attempt to track
only what actually occurred within a single game.
Phrased differently, after the season, we can use the GCP to look backwards and
see how a player performed (just as financial analysts look backwards on
historical quarterly earnings to see how a company performed).

{\footnotesize
\begin{table}[!t]
\centering
{\footnotesize
	\begin{tabular}{cccc}
	Field & Description & \texttt{nba.com} Statistic & \texttt{nba.com} Type\\
	\hline
	MIN & Minutes Played & \texttt{MIN} & Traditional\\
	FG2O & 2 Point Field Goals Made & \texttt{FGM} - \texttt{FG3M} & Traditional\\
	FG2X & 2 Point Field Goals Missed & (\texttt{FGA} - \texttt{FG3A}) -
	FG2O & Traditional\\
	FG3O & 3 Point Field Goals Made & \texttt{FG3M} & Traditional\\
	FG3X & 3 Point Field Goals Missed & \texttt{FG3A} - \texttt{FG3M} & Traditional\\
	FTO & Free Throws Made & \texttt{FTM} & Traditional\\
	FTX & Free Throws Missed & \texttt{FTA} - \texttt{FTM} & Traditional\\
	PF & Personal Fouls & \texttt{PF} & Traditional\\
	STL & Steals & \texttt{STL} & Traditional\\
	BLK & Blocks & \texttt{BLK} & Traditional\\
	TOV & Turnovers & \texttt{TOV} & Traditional\\
	BLKA & Blocks Against & \texttt{BLKA} & Traditional\\
	PFD & Personal Fouls Drawn & \texttt{PFD} & Traditional\\
	POSS & Possessions Played & \texttt{Poss} & Playtype\\
	SAST & Screen Assists & \texttt{SAST} & Hustle\\
	DEFL & Deflections & \texttt{Deflections} & Hustle\\
	CHGD & Charges Drawn & \texttt{Charges Drawn} & Hustle\\
	AC2P & Adj.\ Contested 2PT Shots Defensive &
	\texttt{Contested 2PT Shots} - \texttt{BLK} & Hustle\\
	C3PT & Contested 3PT Shots Defensive &
	\texttt{Contested 3PT Shots} & Hustle\\
	OBOX & Offensive Box Outs & \texttt{OFF BOX OUTS} & Box Outs\\
	DBOX & Defensive Box Outs & \texttt{DEF BOX OUTS} & Box Outs\\
	OLBR & Offensive Loose Balls Recovered &
	\texttt{Off Loose Balls Recovered} & Hustle\\
	DLBR & Defensive Loose Balls Recovered &
	\texttt{Def Loose Balls Recovered} & Hustle\\
	DFGO & Defended Field Goals Made & \texttt{DFGM} & Defensive\\
	DFGX & Defended Field Goals Missed & \texttt{DFGA} - \texttt{DFGM} & Defensive\\
	DRV & Drives & \texttt{Drives} & Tracking\\
	ODIS & Distance Miles Offense & \texttt{Dist.\ Miles Off} & Tracking\\
	DDIS & Distance Miles Defense & \texttt{Dist.\ Miles Def} & Tracking\\
	TCH & Touches & \texttt{Touches} & Tracking\\
	APM & Passes Made & \texttt{Passes Made} -  2AST - PAST
	& Tracking\\
	PASR & Passes Received & \texttt{Passes Received} & Tracking\\
	AST2 & Secondary Assist & \texttt{Secondary Assist} & Tracking\\
	PAST & Potential Assists & \texttt{Potential Assists} & Tracking\\
	OCRB & Contested Offensive Rebounds & \texttt{Contested OREB} & Rebounding\\
	AORC & Adj.\ Offensive Rebound Chances &
	\texttt{OREB Chances} - ORCO & Rebounding\\
	DCRB & Contested Defensive Rebounds & \texttt{Contested DREB} & Rebounding\\
	ADRC & Adj.\ Defensive Rebound Chances &
	\texttt{DREB Chances} - DRCO & Rebounding
	\end{tabular}
}
\caption{
{\small
\textbf{Complete list of statistics used to compute game contribution percentage}.
The statistical categories used to compute the GCP
are listed in the {\it Field} column.  The
{\it Description} column provides a brief description of the statistic in
words.  The statistics in the \texttt{nba.com} {\it Statistic} column are the
source statistics from the \citet{nba_stats}, with formulas as appropriate.
The column \texttt{nba.com} {\it Type} lists the type of statistic used in
terms of the \texttt{nba.com} statistical categories. For complete definitions
and categories of \texttt{nba.com} statistics, please see
\citet{nba_glossary}.}
}
\label{tab:fields}
\end{table}
}

The complete set of fields, $\mathcal{F}$, used in the GCP may be found in
Table~\ref{tab:fields}, along with descriptions, adjustment formulas, and
references to \texttt{nba.com} statistics \citep{nba_stats}.  The fields in
Table~\ref{tab:fields} are meant to align with the principles just outlined.
Nonetheless, we certainly concede alternative choices may be preferable to
other analysts.  Indeed, it may be a collaborative effort between coaches,
scouts, and quantitative departments to determine $\mathcal{F}$.  For our
purposes, we proceed with the 37 fields defined in Table~\ref{tab:fields}.
For a discussion of potential future customization of the GCP, see
Section~\ref{sec:disc}.

Once $\mathcal{F}$ has been
determined, the GCP calculation proceeds as follows.  Let
$g \in \mathcal{G} \equiv \{1, \ldots, 1230\}$ be one of the 1{,}230 games
played in a standard 82-game NBA regular season (recall we assume an investment
horizon of the regular season as an illustration; this may be changed without
materially changing our framework).
Each game, $g$, will consist
of two teams, $t_1, t_2 \in \mathcal{T}$, where $t_1 \neq t_2$ and
$\mathcal{T} \equiv \{\texttt{ATL}, \ldots, \texttt{WAS}\}$ is the set of 30
NBA teams.  Each $t_i$, $i = 1, 2$, will consist of a set of the game's active
players, $\mathcal{P}_{t_i}^g$.
We desire to calculate a GCP per player, per team.
Formally, for $g \in \mathcal{G}$, $t_i \in \mathcal{T}$, and
$p \in \mathcal{P}^g_{t_i}$,
\begin{equation}
\text{GCP}_{t_i, p}^{g} =
\omega_{t_i}^g
\sum_{f \in \mathcal{F}^0_{t_i}} \frac{f_p}{f_{t_i}},
\label{eq:GCP_calc}
\end{equation}
where $f_p$ is the game value of field $f \in \mathcal{F}^0_{t_i}$ for player
$p \in \mathcal{P}^g_{t_i}$, $f_{t_i}$ is team $t_i$'s game total for field
$f \in \mathcal{F}^0_{t_i}$ or
\begin{equation}
f_{t_i} = \sum_{p \in \mathcal{P}^g_{t_i}} f_p,
\label{eq:fti}
\end{equation}
$\mathcal{F}^0_{t_i}$ is the set of fields such that the game totals
$f_{t_i} > 0$, i.e.,
$\mathcal{F}^0_{t_i} = \{f \in \mathcal{F}: f_{t_i} > 0\}$, and
\begin{equation}
\omega^g_{t_i} = \frac{1}{ \mathbf{card} \{ \mathcal{F}^0_{t_i} \} }.
\label{eq:omega}
\end{equation}
Restricting the calculation of $\omega_{t_i}^g$ to only those fields with a
positive team value explicitly ignores any missed categories in the GCP
calculation.  In this way, \eqref{eq:GCP_calc} is dynamic and dependent on a
team's performance.  We acknowledge alternative approaches may be preferable.
For example, it may be desirable to keep the fields and weights fixed, which
would imply that a team recording no instances of a particular field would be
a loss of credit for all players.  We expand on the important choice of
weights and possible future iterations in Section~\ref{sec:disc}.
The calculation in \eqref{eq:GCP_calc} is calculated for each team, i.e., for
$i = 1, 2$.

There are some instructive properties of $\text{GCP}_{t_i, p}^{g}$,
which we now review.  First, for all $g \in \mathcal{G}$, $i = 1, 2$.
\begin{equation}
\sum_{p \in \mathcal{P}^g_{t_i}} \text{GCP}_{t_i, p}^{g} =
\omega_{t_i}^g \sum_{p \in \mathcal{P}^g_{t_i}}
\sum_{f \in \mathcal{F}^0_{t_i}} \frac{f_p}{f_{t_i}} = 
\omega_{t_i}^g \sum_{f \in \mathcal{F}^0_{t_i}} \frac{1}{f_{t_i}}
\sum_{p \in \mathcal{P}^g_{t_i}} f_p
= 1,
\label{eq:GCP_1}
\end{equation}
by \eqref{eq:fti} and \eqref{eq:omega}.  Thus, the sum total of each player's
GCP for each team will be unity for every game.  This makes direct comparisons
possible, and it does not require standardization, such as a need to report
metrics {\it per 100 possessions} \citep{nba_leaders}.
Second, because of NBA forfeit rules and the
statistical categories {\it minutes} and {\it possessions} played
may be recorded by a
player without touching the ball or even moving, the upper bound of GCP is
less than unity for a single player, $p$,
\begin{equation*}
0 \leq \text{GCP}_{t_i, p}^{g} \leq
1 - \omega_{t_i}^g \bigg[
\frac{\text{MIN}_{t_i} - \text{MIN}_p}{\text{MIN}_{t_i}} +
\frac{\text{POSS}_{t_i} - \text{POSS}_p}{\text{POSS}_{t_i}} \bigg],
\quad \text{for all } g \in \mathcal{G}, i = 1, 2.
\end{equation*}
Finally, we emphasize \eqref{eq:GCP_1} holds for both the
winning and losing team.  We refer again to the principles
{\it value all activity} and {\it sign and affect agnostic} as justification
for why each team summing to unity regardless of the team's win-loss outcome is
a desirable property.

\subsubsection{Illustrative Calculation}

For the purposes of illustration, we will consider the April 4, 2023 game
between the \texttt{Philadelphia 76ers} and the \texttt{Boston Celtics}. The
\texttt{76ers} won the game 103-101.  It was a notable game because Joel
Embiid scored 52 points for the \texttt{76ers}, and the game was televised
nationally in the United States on TNT \citep{nba_schedule}.  The game
statistics corresponding to the fields in Table~\ref{tab:fields}, and GCP
calculations for \texttt{Boston} and \texttt{Philadelphia} may be found in Tables~\ref{tab:ill_BOS} and \ref{tab:ill_PHI}, respectively.

The high player for \texttt{Boston} was Jayson Tatum, with a GCP of 20.64\%.
If we consider that Tatum's 37.8 minutes represent only 15.75\% of the total
240 minutes, we can see that Tatum has an out-sized impact on the game in
comparison to a basic minutes played percentage calculation.
The next two players for \texttt{Boston} are Derrick White and Marcus Smart,
with GCPs of 14.25\% and 14.02\%, respectively.  Not close behind are Al
Horford and Malcolm Brogdon, at 13.20\% and 12.43\%, respectively.  These
results suggest \texttt{Boston} had a fairly balanced contribution in this
game.  It is also interesting to observe that Grant Williams had a GCP of
6.34\% in 28.7 minutes, whereas Luke Kornet recorded a higher GCP of 9.12\%
in 15.6 minutes.  Because Kornet contributed more to the game in less playing
time according to \eqref{eq:GCP_calc}, it is a sign that GCP may offer insights
into team building or game management for player personnel officials within
basketball organizations.  For reference, \texttt{Boston} did not record a
CHGD or DLBR in this game, so the categorical weight for \texttt{Boston},
\eqref{eq:omega}, was $\frac{1}{35}$.

For \texttt{Philadelphia}, we see that Joel Embiid recorded a game-high
GCP of 25.30\% in 38.6 minutes.  Based on the histogram of non-zero
GCPs for all players in the 2022-2023 NBA regular season
(i.e., Figure~\ref{fig:gcp_hist}), we see that Joel Embiid had a
99.54\% percentile non-zero GCP game.
The next highest player for \texttt{Philadelphia} is
James Harden, with a GCP of 21.79\%.  It is interesting to see that
\texttt{Philadelphia} had two players with GCPs over 20\%, whereas
\texttt{Boston} had only one in Tatum.  Indeed, \texttt{Philadelphia} had only
two more players above a 10\% GCP, in Tobias Harris and Tyrese Maxey, at
11.86\% and 11.53\%, respectively.  In comparison with \texttt{Boston}, we can
see that \texttt{Philadelphia} was more reliant on less players than
\texttt{Boston}.  This again illustrates some of the added insights of the
GCP metric.  For reference, \texttt{Philadelphia} did not record a OBOX in this
game, so the categorical weight for \texttt{Philadelphia}, \eqref{eq:omega},
was $\frac{1}{36}$.

\begingroup

\setlength{\tabcolsep}{3pt} 
\renewcommand*\rot[2]{\multicolumn{1}{R{#1}{#2}}}

{\normalsize
\begin{table}[!t]
\centering
{\footnotesize
	\begin{tabular}{ccccccccccccc}
& Tatum & Williams & Horford & Smart & White & Brogdon & Hauser & Kornet & Muscala & Griffin\\
\hline
MIN & 37.8 & 28.7 & 34.6 & 30.2 & 40.4 & 27.7 & 3.3 & 15.6 & 13.4 & 8.3\\
FG2O & 5 & 2 & 1 & 5 & 5 & 5 & 0 & 0 & 0 & 0\\
FG2X & 7 & 1 & 1 & 3 & 3 & 7 & 0 & 1 & 0 & 0\\
FG3O & 2 & 2 & 3 & 2 & 4 & 2 & 0 & 0 & 0 & 0\\
FG3X & 6 & 2 & 7 & 5 & 6 & 2 & 1 & 0 & 1 & 0\\
FTMO & 3 & 0 & 0 & 1 & 4 & 2 & 0 & 0 & 0 & 0\\
FTMX & 2 & 0 & 0 & 2 & 0 & 2 & 0 & 0 & 0 & 0\\
PF & 2 & 3 & 4 & 4 & 3 & 0 & 0 & 0 & 0 & 1\\
STL & 3 & 0 & 0 & 1 & 0 & 0 & 0 & 0 & 0 & 0\\
BLK & 0 & 0 & 0 & 0 & 2 & 0 & 0 & 1 & 0 & 1\\
TOV & 2 & 0 & 0 & 3 & 2 & 1 & 0 & 0 & 0 & 0\\
BLKA & 4 & 0 & 0 & 0 & 0 & 2 & 0 & 1 & 0 & 0\\
PFD & 4 & 2 & 0 & 5 & 4 & 5 & 0 & 1 & 0 & 0\\
POSS & 72 & 53 & 65 & 58 & 74 & 52 & 9 & 30 & 25 & 12\\
SAST & 0 & 0 & 3 & 1 & 1 & 0 & 0 & 2 & 1 & 0\\
DEFL & 2 & 0 & 0 & 3 & 0 & 0 & 0 & 0 & 0 & 0\\
CHGD & 0 & 0 & 0 & 0 & 0 & 0 & 0 & 0 & 0 & 0\\
AC2P & 0 & 5 & 10 & 0 & 3 & 2 & 0 & 7 & 3 & 0\\
C3P & 4 & 0 & 5 & 2 & 2 & 2 & 0 & 3 & 0 & 0\\
OBOX & 0 & 0 & 0 & 0 & 0 & 0 & 0 & 1 & 0 & 0\\
DBOX & 0 & 0 & 1 & 0 & 0 & 0 & 0 & 0 & 1 & 1\\
OLBR & 2 & 0 & 1 & 0 & 0 & 1 & 0 & 0 & 0 & 1\\
DLBR & 0 & 0 & 0 & 0 & 0 & 0 & 0 & 0 & 0 & 0\\
DFGO & 3 & 10 & 11 & 6 & 5 & 4 & 1 & 5 & 2 & 5\\
DFGX & 4 & 4 & 8 & 3 & 6 & 9 & 0 & 4 & 3 & 1\\
DRV & 10 & 3 & 1 & 10 & 9 & 14 & 0 & 0 & 0 & 0\\
ODIS & 1.4 & 1.0 & 1.2 & 1.1 & 1.5 & 1.0 & 0.2 & 0.6 & 0.5 & 0.2\\
DDIS & 1.0 & 0.8 & 0.9 & 0.9 & 1.2 & 0.8 & 0.1 & 0.5 & 0.5 & 0.2\\
TCH & 73 & 25 & 56 & 66 & 70 & 52 & 1 & 8 & 9 & 10\\
APM & 32 & 13 & 31 & 35 & 40 & 21 & 0 & 6 & 7 & 10\\
PASR & 52 & 17 & 32 & 52 & 49 & 40 & 1 & 3 & 1 & 4\\
AST2 & 2 & 0 & 0 & 0 & 1 & 1 & 0 & 0 & 0 & 0\\
PAST & 14 & 4 & 11 & 9 & 7 & 7 & 0 & 0 & 0 & 0\\
OCRB & 0 & 1 & 1 & 0 & 0 & 0 & 0 & 2 & 0 & 1\\
AORC & 9 & 2 & 3 & 3 & 1 & 2 & 0 & 2 & 0 & 3\\
DCRB & 0 & 0 & 1 & 0 & 0 & 0 & 0 & 0 & 1 & 2\\
ADRC & 6 & 3 & 6 & 5 & 8 & 6 & 0 & 2 & 3 & 2\\
\hline
GCP & 0.2064 & 0.0634 & 0.1320 & 0.1402 & 0.1425 & 0.1243 & 0.0037 & 0.0912 & 0.0380 & 0.0582
	\end{tabular}
}
\caption{
{\small
\textbf{Game contribution percentage illustration: Boston Celtics}.
The statistical totals for the April 4, 2023 game between the
\texttt{Philadelphia 76ers} and the \texttt{Boston Celtics} for
\texttt{Boston} and ultimate GCP calculation using
\eqref{eq:GCP_calc}.
Players with no recorded statistics or inactive players are not reported. The
statistics were pulled from \citet{nba_stats}. The field abbreviations may
be found in Table~\ref{tab:fields}.}
}
\label{tab:ill_BOS}
\end{table}
}
\endgroup

\begingroup

\setlength{\tabcolsep}{3pt} 
\renewcommand*\rot[2]{\multicolumn{1}{R{#1}{#2}}}

{\normalsize
\begin{table}[!t]
\centering
{\footnotesize
	\begin{tabular}{ccccccccccccccc}
Field & Harris & Tucker & Embiid & Maxey & Harden & Melton & Niang & McDaniels & House Jr. & Reed\\
\hline
MIN & 34.43 & 27.45 & 38.6 & 39.63 & 40.02 & 19.17 & 15.52 & 15.1 & 0.83 & 9.25\\
FG2O & 1 & 1 & 20 & 1 & 3 & 0 & 0 & 1 & 0 & 1\\
FG2X & 3 & 1 & 4 & 3 & 5 & 1 & 1 & 1 & 0 & 2\\
FG3O & 1 & 3 & 0 & 1 & 4 & 0 & 0 & 2 & 0 & 0\\
FG3X & 3 & 0 & 1 & 3 & 5 & 3 & 2 & 1 & 0 & 0\\
FTMO & 0 & 0 & 12 & 0 & 2 & 0 & 0 & 0 & 0 & 0\\
FTMX & 0 & 0 & 1 & 0 & 0 & 0 & 0 & 0 & 0 & 1\\
PF & 4 & 4 & 3 & 5 & 1 & 2 & 0 & 1 & 0 & 1\\
STL & 0 & 0 & 0 & 0 & 1 & 1 & 0 & 0 & 0 & 1\\
BLK & 0 & 0 & 2 & 1 & 2 & 0 & 0 & 1 & 0 & 1\\
TOV & 1 & 2 & 3 & 4 & 0 & 0 & 0 & 0 & 0 & 0\\
BLKA & 1 & 0 & 1 & 0 & 1 & 0 & 0 & 1 & 0 & 0\\
PFD & 0 & 0 & 9 & 2 & 5 & 0 & 0 & 0 & 0 & 1\\
POSS & 64 & 51 & 73 & 75 & 73 & 37 & 29 & 28 & 3 & 15\\
SAST & 1 & 1 & 1 & 0 & 0 & 0 & 0 & 0 & 0 & 1\\
DEFL & 0 & 0 & 0 & 0 & 4 & 1 & 0 & 0 & 0 & 2\\
CHGD & 0 & 0 & 0 & 0 & 1 & 0 & 0 & 0 & 0 & 0\\
AC2P & 1 & 0 & 7 & 2 & 1 & 4 & 0 & 1 & 1 & 2\\
C3P & 1 & 4 & 5 & 2 & 3 & 1 & 1 & 3 & 0 & 1\\
OBOX & 0 & 0 & 0 & 0 & 0 & 0 & 0 & 0 & 0 & 0\\
DBOX & 1 & 1 & 2 & 1 & 0 & 0 & 0 & 0 & 0 & 0\\
OLBR & 0 & 0 & 0 & 0 & 1 & 0 & 1 & 0 & 0 & 0\\
DLBR & 0 & 0 & 1 & 2 & 0 & 0 & 0 & 0 & 0 & 0\\
DFGO & 3 & 4 & 12 & 3 & 7 & 7 & 2 & 4 & 0 & 1\\
DFGX & 6 & 6 & 13 & 7 & 9 & 9 & 2 & 2 & 1 & 3\\
DRV & 1 & 0 & 10 & 3 & 17 & 1 & 1 & 1 & 0 & 1\\
ODIS & 1.2 & 1.0 & 1.2 & 1.4 & 1.3 & 0.7 & 0.6 & 0.6 & 0.0 & 0.4\\
DDIS & 1.1 & 0.9 & 1.2 & 1.4 & 1.1 & 0.6 & 0.5 & 0.5 & 0.1 & 0.3\\
TCH & 31 & 19 & 75 & 62 & 100 & 16 & 17 & 19 & 0 & 8\\
APM & 21 & 14 & 33 & 48 & 63 & 11 & 14 & 12 & 0 & 4\\
PASR & 14 & 13 & 59 & 49 & 83 & 10 & 9 & 9 & 0 & 3\\
AST2 & 1 & 0 & 0 & 0 & 0 & 0 & 0 & 0 & 0 & 0\\
PAST & 1 & 0 & 7 & 1 & 15 & 1 & 0 & 2 & 0 & 1\\
OCRB & 0 & 0 & 1 & 0 & 0 & 0 & 0 & 0 & 0 & 1\\
AORC & 0 & 1 & 2 & 0 & 2 & 0 & 1 & 1 & 0 & 4\\
DCRB & 3 & 0 & 2 & 0 & 0 & 0 & 0 & 0 & 0 & 0\\
ADRC & 5 & 3 & 19 & 9 & 10 & 0 & 1 & 7 & 0 & 1\\
\hline
GCP & 0.1186 & 0.0657 & 0.2530 & 0.1153 & 0.2179 & 0.0524 & 0.0372 & 0.0503 & 0.0026 & 0.0871
	\end{tabular}
}
\caption{
{\small
\textbf{Game contribution percentage illustration: Philadelphia 76ers}.
The statistical totals for the April 4, 2023 game between the
\texttt{Philadelphia 76ers} and the \texttt{Boston Celtics} for
\texttt{Philadelphia} and ultimate GCP calculation using
\eqref{eq:GCP_calc}.
Players with no recorded statistics or inactive players are not reported.
The statistics were pulled from \citet{nba_stats}.
The field abbreviations may be found in Table~\ref{tab:fields}.}
}
\label{tab:ill_PHI}
\end{table}
}
\endgroup

\subsection{Return on Investment}
\label{subsec:ROI_method}

With the GCP methodology sufficiently established, it is now possible to
proceed
to the ROI calculations.  The first step is deriving the SGV in dollars, which
may then be allocated to each player via GCP.  Hence, with a player's salary
serving as a time zero investment to the realized 82-game regular season cash
flows, we are able to employ standard financial
return calculation methods.  In this way, our approach shares some similarity
with the aforementioned \citet{tunaru_2005}, as the number of points recorded
by each player in \citet{tunaru_2005} is translated into a Game Score.
Our GCP methodology differs the Game Score of \citet{tunaru_2005}, however,
and we do not employ an option valuation framework that
relies on a geometric Brownian motion assumption (as mentioned previously,
we also focus on basketball rather than football (i.e., soccer)).
We elaborate on our financial methods as follows.

\subsubsection{Currency Conversion}
\label{subsubsec:currency}

Mechanically, currency conversion calculations are straightforward once the
benchmark items are identified; it is nothing more than a unit conversion.
In this
case, we desire to convert a single NBA game unit into a dollar value of
player compensation using the sum total of player salary.  This differs from
attempting to estimate the dollar value of an NBA game generally, which would
rely on factors such as ticket sales, television revenue, and other items
outside the performance of the players on the court.  To do so, we'll borrow
from the {\it gold standard}, in which a country's basic monetary unit is
defined in terms of the weight of gold {\it specie} \citep{hughes_2011}.
Rather than specie, we will pin the total dollar amount of player compensation
to twice the number of NBA regular season games (i.e., the total number of
single team game units or GCP opportunities).  This is because we are
assuming a regular season investment horizon.  Specifically, there are 1{,}230
NBA regular season games, each of which has two participants within
$\mathcal{T}$.  Hence, we find the SGV through the
direct conversion
\begin{equation}
	\text{SGV} = \frac{ \mathcal{S} }{ 2 \times 1230 },
	\label{eq:SGV}
\end{equation}
where $\mathcal{S}$ is the total dollar value of player compensation for all
players active on NBA rosters for the regular season.  The choices behind
\eqref{eq:SGV} have been made largely for illustrative purposes and have
understandable limitations.  For example, we assign all games the same value,
which may not be appropriate, especially as teams are gradually eliminated
from play-off contention.  Further, to paraphrase a refreshing spin on a
standard NBA pundit platitude regarding the difference between a regular season
and play-off game: there are 82-game players, and there are 16-game players.
(This phrasing is generally attributed to Draymond Green \citep{mahoney_2019}.)
In other words, the best players are paid for play-off games, and so our
regular-season based conversion
would be missing this important component of player on court performance.  We
expand on these points further in Section~\ref{sec:disc}.

\subsubsection{Financial Details}

\begin{figure}[t!]
    \centering
    \includegraphics[scale=1.5]{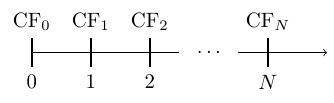}
\caption{
{\small
\textbf{Cash flow time line}.
A classical illustration of a sequence of financial cash flows.  Within the
context of the NBA, we assume $\text{CF}_0$ represents the season salary
and is thus a negative cash flow to the organization.  The remaining cash
flows, $\text{CF}_1$ through $\text{CF}_N$ represent the dollar conversion of
the player's on court production, which may be estimated using
\eqref{eq:GCP_calc} and \eqref{eq:SGV}.  Observe $\text{CF}_i \geq 0$ for
$1 \leq i \leq N$ by \eqref{eq:GCP_calc} and \eqref{eq:SGV}.
Because we assume a traditional NBA regular season for illustration,
$N = 82$ (though this may be adjusted to meet an analyst's needs). For
simplicity, we assume the games are played at equally spaced intervals, though
this assumption may also be relaxed.  See Section~\ref{sec:disc} for further
discussion.}
}
    \label{fig:cf_figure}
\end{figure}

We now briefly review how to calculate the realized return for a sequence of
financial cash flows.  Because we know the initial investment (i.e., a player's
salary) and subsequent cash flows (i.e., a player's vector of GCPs multiplied
by the SGV), we will utilize the {\it internal rate of return} methodology
\citep[e.g.,][\S 4.8]{berk_2007}.  Consider the time line of cash flows
illustrated in Figure~\ref{fig:cf_figure}.  For simplicity, we assumed each
cash flow, $\text{CF}_1, \ldots, \text{CF}_N$, is paid on equally-spaced
intervals.  This assumption may be relaxed, a point we address more fully in
Section~\ref{sec:disc}.  Further, we assumed the time zero cash flow,
$\text{CF}_0$, is the initial investment.  Because we are performing a realized
return calculation (i.e., after the completion of the regular season), all
$\text{CF}_i$, $0 \leq i \leq n$ are known at the onset of the problem. The
return on investment is the rate, $r$, such that
\begin{equation}
\text{CF}_0 = \sum_{i = 1}^{N} \frac{ \text{CF}_i }{ (1 + r)^i }.
\label{eq:IRR_calc}
\end{equation}
Aside from very simplified versions of \eqref{eq:IRR_calc}, the computation of
$r$ will require the use of optimization software (we have found the
\texttt{irr} function in the \texttt{R} package \texttt{jrvFinance}
\citep{varma_2021} useful for this purpose).

Let us now interpret \eqref{eq:IRR_calc} within the context of an NBA regular
season.  The time zero cash flow (i.e., the initial investment) is a player's
salary.  From the perspective of the NBA team it is a negative cash flow.
Because we focus on an NBA regular season, it is natural to assume $N = 82$
(though it is possible a player traded during the NBA regular season may
appear in more than 82 games, such as Mikal Bridges in the 2022-2023 NBA
regular season).
To find the return cash flows, $\text{CF}_i$, $1 \leq i \leq 82$, we may use
\eqref{eq:GCP_calc} and \eqref{eq:SGV}.  Formally, let $p^*$ represent an
active NBA player on team $t^* \in \mathcal{T}$ and let
$\bm{g}_* = \{g^*_1, \ldots, g^*_{82}\}$ be the
ordered sequence of games in $\mathcal{G}$ of which $t^*$ appeared.  Then,
\begin{equation}
	\text{CF}_{i, p^*} = \text{SGV} \times \text{GCP}^{g^*_i}_{t^*, p^*}.
	\label{eq:CF_i}
\end{equation}
By definition of \eqref{eq:GCP_calc} and \eqref{eq:SGV}, it is clear
$\text{CF}_{i, p^*} \geq 0$ for all $1 \leq i \leq N$ and $p^*$.  Hence, the
contractual return on investment for $p^*$ is the rate, $r_{p^*}$, such that
\begin{equation}
	\text{CF}_{0, p^*} = \text{SGV} \sum_{i=1}^{N}
	\frac{ \text{GCP}^{g^*_i}_{t^*, p^*} }{ (1 + r_{p^*})^i }.
	\label{eq:NBA_IRR}
\end{equation}
We remark that any player traded during the regular season will require the
standard adjustments to \eqref{eq:CF_i} (we make these adjustments for all
results in Section~\ref{sec:results}).  We also remark briefly that the
framework in \eqref{eq:NBA_IRR} may be adjusted for enhanced precision.  For
example, rather than assume a player's salary is a time zero lump sum payment,
the true timing of salary payments may be incorporated into
\eqref{eq:IRR_calc}.  Similarly, there is no need to assume all games are
played on equally spaced intervals, and the true dates and time calculations
may be made more precise.  Further, we consider only on court performance in
the form of GCP, but it is not unreasonable to credit off court revenue to
players, such as with team jersey sales.  Lastly, $N$ may be extended to also
include play-off games, and the SGV or GCP metrics may be adjusted to fit an
analysts' preference.  As a final comment, the nature of \eqref{eq:NBA_IRR}
naturally assigns more weight to early season performance, which may also be
adjusted by reordering or weighting the games to meet an analysts' needs.
It is the framework we propose that we feel is of
general value.  Further discussion on potential enhancements or customization
may be found in Section~\ref{sec:disc}.

\section{Results}
\label{sec:results}

As an illustration,
we now apply the methods of Section~\ref{sec:methods} to the pool of players
participating the 2022-2023 NBA regular season.  We first briefly discuss the
data and how it was obtained.  Next, we present cumulative sums of complete
regular season GCPs.  This is done to help validate the GCP metric itself and
demonstrate it has utility irrespective of player salary data.  Finally, the
section concludes by presenting ROI calculations, the ultimate purpose of this
study.   In support of reproducible research,
the complete compiled data and replication code may be found on a public
\texttt{github} repository at
\href{https://github.com/jackson-lautier/nba_roi}{https://github.com/jackson-lautier/nba\_roi}.

\subsection{Data}

Our data comes from two publicly available sources.  The first is the
statistics page of the NBA \citep{nba_stats}.  Because of the extensive nature
of the statistical categories ultimately utilized in the GCP calculation, we
found the \texttt{python} package \texttt{nba\_api} \citep{nba_api} of
enormous value.  Indeed, we compiled a novel dataset of box scores of the
form of Tables~\ref{tab:ill_BOS} and \ref{tab:ill_PHI} for all 1{,}230 games of
the 2022-2023 NBA regular season by designing a custom game-by-game query
wrapper for \citet{nba_api}.  The second is a complete list of player salaries
for the 2022-2023 NBA regular season from \citet{hoops_hype} (with one
supplement for the player Chance Comanche \citep{chance_sal}).  Both of these
sources were combined into a novel dataset that includes both on court
game-by-game performance in the form of Tables~\ref{tab:ill_BOS} and
\ref{tab:ill_PHI} and player salary data. To obtain the data and replication
code, please navigate to the public
\texttt{github} repository at
\href{https://github.com/jackson-lautier/nba_roi}{https://github.com/jackson-lautier/nba\_roi}.

\subsection{Absolute Game Contribution Percentage}
\label{subsec:PVGCP_results}

\begin{figure}[t!]
    \includegraphics[width=\textwidth]{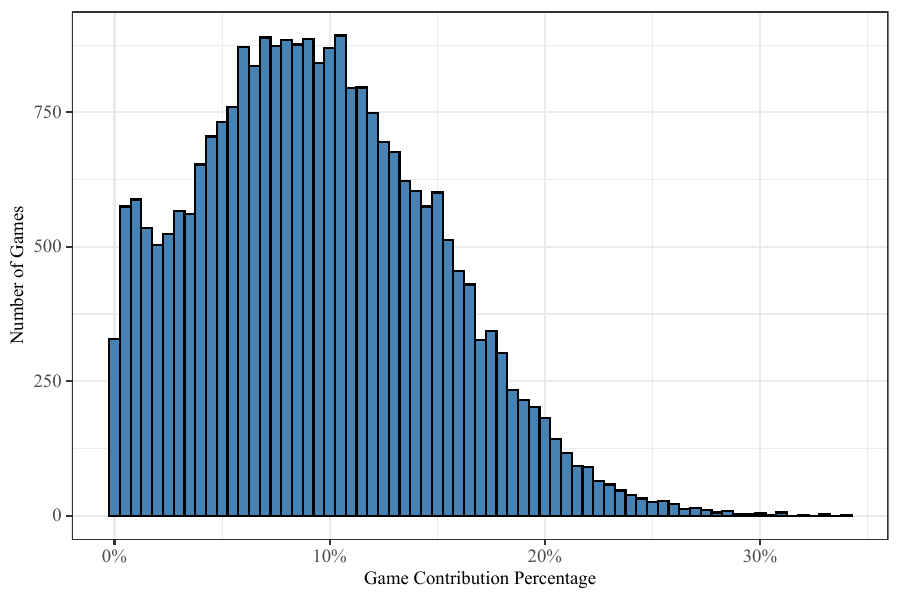}
\caption{
{\small
\textbf{Game contribution percentage histogram}.
A distributional plot of the 25{,}892 non-zero GCP realizations for all
players in the 2022-2023 NBA regular season.  The GCP was computed using the
methods of Section~\ref{subsec:GCP_method}.  For reference, the maximum
realized GCP was by Luka Don\v{c}i\'{c} at 33.8\%, which occurred against the
\texttt{New York Knicks} on December 27, 2023.  This compares favorably to the
30-35\% maximum player salary allowable under the NBA's Collective Bargaining
Agreement \citep{nba_cba}.}
}
    \label{fig:gcp_hist}
\end{figure}

The first set of results is a histogram of all non-zero GCPs recorded for the
2022-2023 NBA regular season, which is available in Figure~\ref{fig:gcp_hist}.
The histogram spans 25{,}892 GCP realizations, and it helps us get a sense of
the distribution of this novel metric.
Specifically, we can see that GCP realizes a
peak near 10\%, with a tail skewed to the right.  The maximum realized GCP was
by Luka Don\v{c}i\'{c} at 33.8\%, which occurred against the
\texttt{New York Knicks} on
December 27, 2023.  The game was notable because Don\v{c}i\'{c} scored 60
points while also recording 21 rebounds and 10 assists.  It was the first
60-point game in the history of the \texttt{Dallas Mavericks}, a career high in
rebounds for Don\v{c}i\'{c}, and, historically, the first 60-20-10 game in NBA
history \citep{espn_dal}.  As a bit of informal verification for the GCP
metric, the maximum player salary ranges from 30-35\% of the salary cap,
subject to years of service and other performance-based criteria
\citep{nba_cba}.  Hence, having a maximum realization of GCP for all players
within the 30-35\% range is in this sense intuitively pleasing.  (We note the
GCP of Section~\ref{subsec:GCP_method} was developed without regard to the
30-35\% maximum salary; it is a satisfying coincidence.)

\begin{figure}[t!]
    \includegraphics[width=\textwidth]{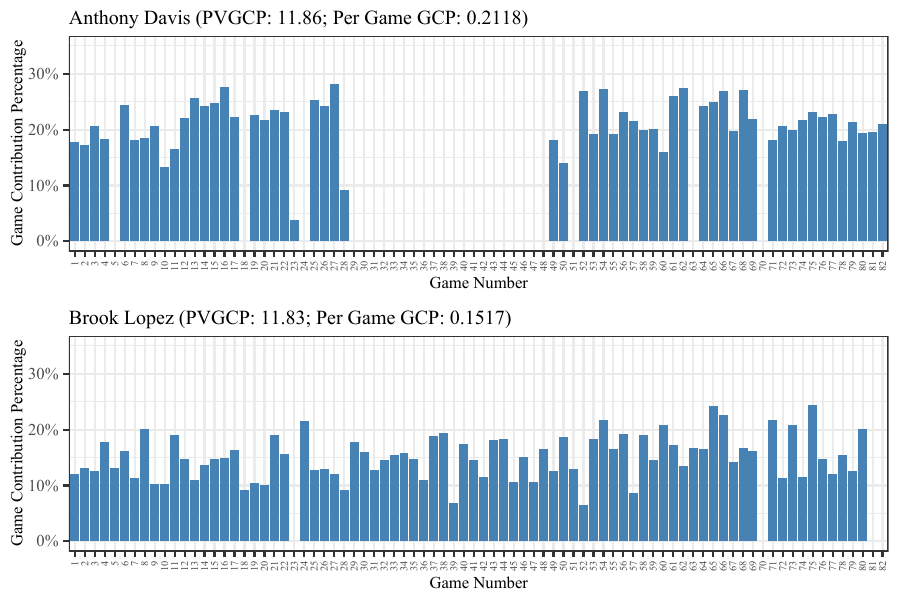}
\caption{
{\small
\textbf{Game contribution percentage comparison}.
A game-by-game plot of GCP for Anthony Davis (top) and Brook Lopez (bottom)
for the 2022-2023 NBA regular season.  Both Davis and Lopez registered nearly
an identical PVGCP (i.e., \eqref{eq:pv_gcp}) at 11.86 and 11.83, respectively,
despite a large difference in games played (56 versus 78, respectively).
Because GCP is a single game calculation, it treats missed games as defaults.
Hence, Davis missed enough games and Lopez played enough games to register a
similar level of cumulative contribution as measured by PVGCP.
Because Lopez earned \$13{,}906{,}976 in comparison to
Davis at \$37{,}980{,}720 for the 2022-2023 NBA regular season \citep{hoops_hype},
this information may be of interest to NBA player personnel decision makers.
}
}
    \label{fig:gcp_comp}
\end{figure}

As a next set of results, we demonstrate how GCP may be used to assess the
tipping point of a player who performs very well but has a tendency to miss
games against a player that performs only reasonably well but does so
consistently.  The difficulty of comparing a high-performing player with many
missed games against an average-performing player with few missed games has
been a frequent source of consternation within the discourse of NBA pundits
(for example, \citet{lowe_allnba} and \citet{beck_2023} frequently cite games
missed as reasoning for preferring some players over others).
Because a player receives a zero GCP for any missed game (see
Figure~\ref{fig:gcp_comp}), we
may find the break even point by taking a running tally of GCP for the period
in question, such as the entire 2022-2023 NBA regular season.  In effect, we
are taking a {\it present value} of all GCPs at an interest rate of 0\%.  That
is, for active NBA player $p^*$ on team $t^* \in \mathcal{T}$ with
$\bm{g}_* = \{g^*_1, \ldots, g^*_{82}\}$ representing the
ordered sequence of games in $\mathcal{G}$ of which $t^*$ appeared, we have
\begin{equation}
	\text{PVGCP}_{p^*} = \sum_{\bm{g}_*} \text{GCP}^{g^*}_{t^*, p^*}.
	\label{eq:pv_gcp}
\end{equation}
As with \eqref{eq:CF_i}, the adjustments to \eqref{eq:pv_gcp} are natural for
any players traded over the time period in question.  We emphasize that
\eqref{eq:pv_gcp} is directly comparable for all NBA players and does not
require standardization, such as per 100 possessions.  This allows for analysts
to use a single metric to understand the impact of a player's missed games,
rather than computing a metric, standardizing it, and then attempting to
perform additional missed game value judgments
\citep[e.g.,][]{lowe_allnba, beck_2023}.

Table~\ref{tab:top50_GCP}
presents the top fifty players in terms of PVGCP for the 2022-2023 NBA regular
season.  We can see the top five performers are Damontas Sabonis, Nikola Joki\'{c},
Joel Embiid, Luka Don\v{c}i\'{c}, and Bam Adebayo.  In general, our PVGCP metric
arrives at a similar list of top performers, as measured by the 2022-2023 Kia
NBA Most Valuable Player award voting \citep{nba_mvp} and 2022-2023 Kia All-NBA
selections \citep{nba_allnba}.  The PVGCP metric prefers Sabonis as the
top performer, whereas Embiid won the Kia NBA Most Valuable Player award.
Table~\ref{tab:top50_GCP} has Embiid ranked 3rd in terms of PVGCP, though
Embiid was the top per game GCP performer.  Further, the PVGCP metric has
Sabonis at a clear top spot, but he received only one 4th place MVP vote and
24 5th place votes.  Hence,
it is not unreasonable to suggest that our GCP methodology
is able to capture Sabonis' consistent high level of contribution to his team's
on court performance and availability in a
way that other popular advanced
metrics may overlook (that said, the win shares approach
\citep{win_share} had Sabonis ranked second behind Joki\'{c} \citep{nba_leaders}).
Aside from win shares, no other Table~\ref{tab:top50_GCP} metric had Sabonis
higher than 7th.
Further, all comparative metrics reported in Table~\ref{tab:top50_GCP} have
Nikola Joki\'{c} as the top performer, which suggests that these popular
advanced metrics may rely on similar overall approaches and not offer enough
diversity of perspective.  For completeness, we note \citet{berri_2010} offers
a critique of sports metrics proposed outside the scope of academic
peer-review.

In this regard, we emphasize
again that GCP is a measure of a player's total contribution to his team's on
court performance; it does not attempt to parse out ``good'' and ``bad''
performance (review Section~\ref{subsec:GCP_method} as needed).  This helps
explain why players like Alperen \c{S}eng\"{u}n and Nikola Vuce\v{c}i\'{c} have high values of
PVGCP despite playing for teams in the \texttt{Houston Rockets} and
\texttt{Chicago Bulls} that amassed paltry win percentages of 0.268 and 0.488,
respectively \citep{nba_stand}.
As a further reflective note, it appears PVGCP is partial
to the traditional {\it big man} in that there is a healthy representation of
centers with high games played that were not recognized on many NBA award
ballots (e.g., Vuce\v{c}i\'{c}, \c{S}eng\"{u}n, Claxton, Gobert, Zubac,
Porzi\c{n}\c{g}is, Valan\v{c}i\={u}nas,
Plumlee, P\"{o}ltl, Looney, and Okongwu).  Then again, the importance of the
center position has been long established in basketball treatments
\citep[e.g.,][pg. 40]{oliver_2004}, and more generally its depth of representation
in Table~\ref{tab:top50_GCP} is a useful insight for NBA player personnel
decision-makers tasked with allocating a capped player salary pool.

\begin{table}[!t]
\centering
{\scriptsize
	\begin{tabular}{cccccccccc}
Rank & Player & GP & PVGCP & GCPpg
& PER & WS & BPM & VORP & RAPTOR\\
\hline
1&Domantas Sabonis&79& \textbf{16.81} &0.213&23.50&12.60&5.80&5.40&8.66\\
2&Nikola Joki\'{c} &69&15.04&0.218& \textbf{31.50} & \textbf{14.90} & \textbf{13.00}
& \textbf{8.80} & \textbf{20.31}\\
3&Joel Embiid&66&14.81& \textbf{0.224} &31.40&12.30&9.20&6.40&12.82\\
4&Luka Don\v{c}i\'{c}&66&14.24&0.216&28.70&10.20&8.90&6.60&12.98\\
5&Bam Adebayo&75&14.10&0.188&20.10&7.40&1.50&2.30&5.69\\
6&Giannis Antetokounmpo&63&13.62&0.216&29.00&8.60&8.50&5.40&9.31\\
7&Evan Mobley&79&13.61&0.172&17.90&8.50&1.70&2.50&3.85\\
8&Nikola Vuce\v{c}i\'{c}&82&13.50&0.165&19.10&8.30&2.70&3.20&1.91\\
9&Julius Randle&77&13.01&0.169&20.30&8.10&3.70&3.90&5.64\\
10&Alperen \c{S}eng\"{u}n&75&12.94&0.173&19.70&5.20&1.40&1.90&5.18\\
11&Jayson Tatum&74&12.88&0.174&23.70&10.50&5.50&5.10&8.99\\
12&Pascal Siakam&71&12.54&0.177&20.30&7.80&3.10&3.40&4.50\\
13&Shai Gilgeous-Alexander&68&12.36&0.182&27.20&11.40&7.30&5.60&9.86\\
14&Anthony Edwards&79&12.34&0.156&17.40&3.80&1.00&2.10&6.42\\
15&Nic Claxton&76&12.25&0.161&20.80&9.20&3.10&2.90&5.57\\
16&Rudy Gobert&70&12.22&0.175&18.90&7.80&0.70&1.40&5.25\\
17&Ivica Zubac&76&12.15&0.160&16.70&6.70&-0.90&0.60&2.15\\
18&Trae Young&73&12.00&0.164&22.00&6.70&3.30&3.40&9.12\\
19&Anthony Davis&56&11.86&0.212&27.80&9.00&6.30&4.00&9.77\\
20&Brook Lopez&78&11.83&0.152&18.40&8.00&2.10&2.50&8.70\\
21&Paolo Banchero&72&11.57&0.161&14.90&2.40&-1.50&0.30&-0.42\\
22&Kristaps Porzi\c{n}\c{g}is&65&11.44&0.176&23.10&7.70&4.30&3.40&8.24\\
23&Scottie Barnes&77&11.35&0.147&15.50&5.00&0.40&1.60&4.62\\
24&Jonas Valan\v{c}i\={u}nas&79&11.34&0.144&19.30&5.80&-0.40&0.80&0.55\\
25&Mason Plumlee&79&11.29&0.143&19.60&7.90&2.20&2.20&3.20\\
26&De'Aaron Fox&73&11.18&0.153&21.80&7.40&2.50&2.70&7.19\\
27&Jalen Brunson&68&11.16&0.164&21.20&8.70&3.90&3.50&8.14\\
28&Jakob P\"{o}ltl&72&11.16&0.155&21.00&6.00&1.90&1.90&3.45\\
29&DeMar DeRozan&74&11.14&0.151&20.60&8.50&2.00&2.60&7.38\\
30&Zach LaVine&77&11.04&0.143&19.00&7.10&1.90&2.70&5.53\\
31&Jarrett Allen&68&11.00&0.162&19.90&9.50&2.40&2.40&4.53\\
32&Fred VanVleet&69&10.94&0.159&17.00&6.50&2.50&2.90&10.01\\
33&Donovan Mitchell&68&10.76&0.158&22.90&8.90&6.30&5.00&9.45\\
34&Mikal Bridges&82&10.73&0.131&16.80&7.50&1.70&2.80&6.76\\
35&Jaylen Brown&67&10.68&0.159&19.10&5.00&1.30&2.00&4.12\\
36&Kevon Looney&82&10.61&0.129&17.80&8.70&2.10&2.00&6.09\\
37&Draymond Green&73&10.59&0.145&12.20&4.70&0.80&1.60&6.09\\
38&Spencer Dinwiddie&79&10.56&0.134&16.00&6.30&0.70&1.80&4.56\\
39&CJ McCollum&75&10.54&0.140&15.60&4.30&0.80&1.90&3.59\\
40&Dejounte Murray&74&10.49&0.142&17.00&4.70&1.00&2.10&3.07\\
41&Jordan Poole&82&10.46&0.128&14.60&3.20&-1.90&0.10&-0.44\\
42&Franz Wagner&80&10.44&0.130&15.90&5.40&-0.10&1.30&7.24\\
43&Jimmy Butler&64&10.40&0.162&27.60&12.30&8.70&5.80&10.11\\
44&Onyeka Okongwu&80&10.39&0.130&19.40&7.10&0.80&1.30&3.50\\
45&Damian Lillard&58&10.34&0.178&26.70&9.00&7.10&4.90&11.52\\
46&Jalen Green&76&10.30&0.136&14.50&1.80&-2.10&0.00&1.75\\
47&Ja Morant&61&10.30&0.169&23.30&6.00&5.70&3.80&8.39\\
48&Russell Westbrook&73&10.30&0.141&16.10&1.90&0.20&1.20&1.37\\
49&Darius Garland&69&10.29&0.149&18.80&7.60&2.40&2.70&8.95\\
50&James Harden&58&10.26&0.177&21.60&8.40&5.40&4.00&9.22\\
\hline
	\end{tabular}
}
\caption{
{\small
\textbf{Top performers: Absolute game contribution percentage}.
The top 50 performers in terms of \eqref{eq:pv_gcp} for the 2022-2023 NBA
regular season.  For reference, we also present the number of games played
(GP), GCP per game (GCPpg), standard player efficiency rating (PER)
\citep{nba_leaders}, win shares (WS) \citep{nba_leaders}, box plus/minus
(BPM) per 100 possessions \citep{nba_leaders}, value over replacement
player (VORP) \citep{nba_leaders}, and regular season wins above replacement
(RAPTOR) \citep{raptor} for each player.  The high value in each column has
been noted in bold.}
}
\label{tab:top50_GCP}
\end{table}

It is interesting to observe that there are multiple ways to obtain an
impressive PVGCP.  As we've established, it is clear that players
with any missed games will be directly penalized in \eqref{eq:pv_gcp} in a way
that differs from metrics based on cumulative on court statistics.
Nonetheless, it is possible for a player to perform so well in games played
that they can amass a high PVGCP despite accumulating many missed games.  With
PVGCP, we can obtain this exact inflection point.  This
is the value of treating missed games as defaults, and it may offer useful
insights on its own merit.  Consider Figure~\ref{fig:gcp_comp}, which compares
Anthony Davis and Brook Lopez.  From Table~\ref{tab:top50_GCP}, we can see that
Davis and Lopez accumulated nearly identical PVGCPs at 11.86 and 11.83,
respectively.  Davis did so in 56 games (i.e., 26 defaults) whereas Lopez did
so in 78 games (i.e., only 4 defaults).  The visual representation in
Figure~\ref{fig:gcp_comp} makes the difference in consistency readily apparent.
In other words, Lopez, through his consistent availability and steady
performance in 78 games was able to reach the same level of PVGCP as Davis for
the 2022-2023 NBA season.  Because Lopez earned \$13{,}906{,}976 in comparison to
Davis at \$37{,}980{,}720 for the 2022-2023 NBA season \citep{hoops_hype}, this
information may be of interest to NBA player personnel decision makers. (Davis
had a very strong 2023 NBA postseason, which is an important consideration not made
within this analysis.)  We arrive at the formal ROI calculations in
Section~\ref{subsec:ROI_results}.

\subsection{Return on Investment}
\label{subsec:ROI_results}

The final component of our effort and the ultimate purpose
of this study is to combine the absolute GCP results of
Table~\ref{tab:top50_GCP} with each player's salary to perform a contractual
realized ROI calculation in the form of \eqref{eq:NBA_IRR}.
The first step is to calculate the SGV proposed in \eqref{eq:SGV}.
Based on
our data of 547 active NBA players for the 2022-2023 NBA season, we have
$\mathcal{S} =$ \$4{,}472{,}678{,}188. Thus, $\text{SGV} =$ \$1{,}818{,}162.
To avoid skewed calculations from players on 10-day contracts, we set a
minimum games played requirement of 25 games.  This limits the calculation
pool to 423 players.  The top and bottom 50 performers in terms of ROI,
$r$, via \eqref{eq:NBA_IRR}, have been compiled in
Tables~\ref{tab:top50_ROI} and \ref{tab:bottom50_ROI}, respectively.  There are
some interesting observations.

We begin with the top performers.  Immediately, we see that the highest salary
in Table~\ref{tab:top50_ROI} is \$4.215M, which belongs to Tyrese Haliburton.
Because this salary is well below the mean (median) player salary of 
\$10.022M (\$5.122M) for all players playing at least 25 games during the
2022-2023 NBA regular season, it is clear that the biggest returns belong to
players signed to small value contracts that also contribute meaningfully in
terms of on court production.  In other words, the ROI calculation prefers
players with a small
initial investment (i.e., $\text{CF}_0$ in Figure~\ref{fig:cf_figure}) that
produce valuable game-by-game output (i.e., $\text{CF}_i$, $1 \leq i \leq 82$
in Figure~\ref{fig:cf_figure}).  This insight is perhaps elementary, but the
methods we propose lead to a direct framework to identify which players have
outperformed their salaries and by how much, neither of which is a
straightforward calculation. If we combine this information
with the NBA team salary cap restrictions, then it may be used to identify
market inefficiencies in hopes of optimizing team roster construction.  On the
player side, this information may be used in upcoming contract negotiations.
(It is notable that a number of players in Table~\ref{tab:top50_ROI} have
signed contracts with large raises for the upcoming 2023-2024 NBA season:
e.g., Tre Jones, Max Strus, Austin Reaves, Jock Landale, Desmond Bane,
Gabe Vincent, Jordan Poole, Shake Milton, Tyrese Haliburton \citep{nba_FA}).
Additionally, it is of interest to observe that there is limited overlap with
Table~\ref{tab:top50_GCP}.  Indeed, only Alperen \c{S}eng\"{u}n and
Jordan Poole appear in both tables.  This again highlights the importance of
the initial investment in calculating an ROI with
\eqref{eq:NBA_IRR}.

\begin{table}[!t]
\centering
{\scriptsize
	\begin{tabular}{cccccc}
Rank & Player & Salary & GP & PVGCP & ROI (\%)\\
\hline
1&Tre Jones& \$1.783&68&8.228&0.132\\
2&Kevon Harris& \$0.509&34&1.778&0.122\\
3&Nick Richards& \$1.783&65&6.766&0.113\\
4&Ayo Dosunmu& \$1.564&80&6.921&0.108\\
5&Max Strus& \$1.816&80&7.336&0.106\\
6&Anthony Lamb& \$0.695&62&4.734&0.096\\
7&Christian Koloko& \$1.500&58&3.851&0.095\\
8&Austin Reaves& \$1.564&64&6.594&0.094\\
9&Jock Landale& \$1.564&69&5.546&0.094\\
10&Jose Alvarado& \$1.564&61&5.475&0.092\\
11&Jaden McDaniels& \$2.161&79&8.587&0.087\\
12&Daniel Gafford& \$1.931&78&9.111&0.086\\
13&Kevin Porter Jr.& \$3.218&59&8.968&0.086\\
14&Kenyon Martin Jr.& \$1.783&82&7.208&0.086\\
15&Santi Aldama& \$2.094&77&6.385&0.080\\
16&Desmond Bane& \$2.130&58&7.823&0.079\\
17&Bol Bol& \$2.200&70&5.524&0.077\\
18&Alperen \c{S}eng\"{u}n& \$3.375&75&12.943&0.077\\
19&Drew Eubanks& \$1.968&78&8.166&0.077\\
20&Herbert Jones& \$1.785&66&7.614&0.076\\
21&Jordan Goodwin& \$1.280&62&5.189&0.075\\
22&Naji Marshall& \$1.783&77&6.417&0.073\\
23&Immanuel Quickley& \$2.316&81&8.902&0.073\\
24&Gabe Vincent& \$1.816&68&6.048&0.073\\
25&Tyrese Maxey& \$2.727&60&7.274&0.073\\
26&Dennis Smith Jr.& \$2.133&54&5.776&0.068\\
27&Jaylen Nowell& \$1.931&65&4.538&0.067\\
28&Terance Mann& \$1.931&81&6.528&0.066\\
29&Kenrich Williams& \$2.000&53&5.004&0.065\\
30&Orlando Robinson& \$0.386&31&2.111&0.063\\
31&Aaron Wiggins& \$1.564&70&4.695&0.062\\
32&Naz Reid& \$1.931&68&6.862&0.060\\
33&Troy Brown Jr.& \$1.968&76&5.707&0.060\\
34&Isaiah Stewart& \$3.433&50&6.489&0.059\\
35&Jeremiah Robinson-Earl& \$2.000&43&3.277&0.059\\
36&Walker Kessler& \$2.696&74&9.751&0.059\\
37&Duane Washington Jr.& \$0.629&31&1.735&0.058\\
38&Wenyen Gabriel& \$1.879&68&5.852&0.057\\
39&John Konchar& \$2.300&72&5.006&0.057\\
40&Jordan Poole& \$3.901&82&10.461&0.056\\
41&Shake Milton& \$1.998&76&5.524&0.056\\
42&Andrew Nembhard& \$2.244&75&6.998&0.056\\
43&Damion Lee& \$2.133&74&4.725&0.054\\
44&Isaiah Jackson& \$2.574&63&5.796&0.054\\
45&Isaiah Livers& \$1.564&52&3.818&0.054\\
46&Keldon Johnson& \$3.873&63&8.391&0.053\\
47&Tyrese Haliburton& \$4.215&56&7.964&0.053\\
48&Javonte Green& \$1.816&32&2.074&0.052\\
49&Trendon Watford& \$1.564&62&5.130&0.051\\
50&Jericho Sims& \$1.640&52&3.731&0.049\\
\hline
	\end{tabular}
}
\caption{
{\small
\textbf{Top 50 performers: Return on investment}.
The top 50 performers in terms of \eqref{eq:NBA_IRR} for the 2022-2023 NBA
regular season, based on a minimum of 25 games played.
For reference, we also include the player salary \citep{hoops_hype} in
millions, games played (GP), and PVGCP, as calculated with
\eqref{eq:pv_gcp}.}
}
\label{tab:top50_ROI}
\end{table}

The lowest returns in Table~\ref{tab:bottom50_ROI} offer another set of
interesting observations.  As expected, there are many large contracts in
Table~\ref{tab:bottom50_ROI}, many of which are well above the 75th percentile
salary of \$13.64M for all players playing at least 25 games during the
2022-2023 NBA regular season.  As the opposite reasoning of the previous
paragraph would suggest, it requires much stronger on court performance
(i.e., $\text{CF}_i$, $1 \leq i \leq 82$) to overcome a much higher initial
investment (i.e., $\text{CF}_0$).  In addition, there are many highly
decorated NBA players in Table~\ref{tab:bottom50_ROI}, such as Stephen Curry,
LeBron James, Kevin Durant, and Giannis Antetokounmpo.
This may be surprising at first glance, but we offer a few reasonable
explanations.  First, many players in Table~\ref{tab:bottom50_ROI}
missed games in the 2022-2023 NBA regular season.  Quite simply, it is
difficult to overcome a large initial investment with many subsequent zero
cash flows.  Second, we do not include playoff games in the calculations for
Table~\ref{tab:bottom50_ROI}.  If NBA personnel decision makers put a premium
on playoff performance (a very reasonable supposition), then the calculations
in Table~\ref{tab:bottom50_ROI} are missing an important component of the
contractual value of the highest paid NBA players.  Similarly, we only consider
on court performance, and we ignore off court value vis-\'{a}-vis jersey sales,
ticket sales, television revenue, and other potential pecuniary production that
is a likely income component to teams rostering the NBA's most popular
players.  We attempted to value on court performance only by design,
but this is a straightforward adjustment to the ROI framework we propose.
Additional related discussion may be found in Section~\ref{sec:disc}.  As a
final reference point, the highest 2022-2023 salary was \$48.07, which belonged
to Stephen Curry.  Assuming all 82 games played, the break-even IRR implies a
per game cash flow of \$0.586M.  Assuming an SGV of \$1.818M, as calculated
above, this implies a per game GCP of 32.24\%.  Again, this is quite close to
the maximum player salary of 30-35\% per the NBA's CBA \citep{nba_cba} and is
an additional informal validation of our approach.

\begin{table}[!t]
\centering
{\scriptsize
	\begin{tabular}{cccccc}
Rank & Player & Salary & GP & PVGCP & ROI (\%)\\
\hline
423&Derrick Rose& \$14.521&27&1.23&-0.080\\
422&John Wall& \$47.346&34&3.51&-0.070\\
421&Evan Fournier& \$18.000&27&1.33&-0.047\\
420&Andrew Wiggins& \$33.617&37&4.76&-0.039\\
419&Ben Simmons& \$35.449&42&5.33&-0.038\\
418&Garrett Temple& \$5.156&25&0.52&-0.037\\
417&Duncan Robinson& \$16.902&42&2.10&-0.032\\
416&Richaun Holmes& \$11.215&42&1.53&-0.032\\
415&Karl-Anthony Towns& \$33.833&29&4.58&-0.031\\
414&Davis Bertans& \$16.000&45&1.62&-0.030\\
413&Khris Middleton& \$37.984&33&3.63&-0.029\\
412&Bradley Beal& \$43.279&50&7.11&-0.026\\
411&Stephen Curry& \$48.070&56&8.72&-0.023\\
410&LeBron James& \$44.475&55&9.15&-0.022\\
409&Kyle Lowry& \$28.333&55&6.81&-0.022\\
408&Zion Williamson& \$13.535&29&4.67&-0.022\\
407&Klay Thompson& \$40.600&69&7.26&-0.022\\
406&Gordon Hayward& \$30.075&50&5.49&-0.022\\
405&Paul George& \$42.492&56&9.24&-0.021\\
404&Tobias Harris& \$37.633&74&8.33&-0.020\\
403&Bryn Forbes& \$2.298&25&0.71&-0.020\\
402&Michael Porter Jr.& \$30.914&62&6.29&-0.020\\
401&Collin Sexton& \$16.700&48&4.66&-0.020\\
400&Wendell Moore Jr.& \$2.307&29&0.53&-0.020\\
399&Furkan Korkmaz& \$5.000&37&1.13&-0.020\\
398&Kawhi Leonard& \$42.492&52&7.98&-0.019\\
397&Joe Harris& \$18.643&74&4.85&-0.019\\
396&Boban Marjanovic& \$4.101&31&0.80&-0.019\\
395&Damian Lillard& \$42.492&58&10.34&-0.018\\
394&Patty Mills& \$6.479&40&1.68&-0.018\\
393&Matthew Dellavedova& \$2.629&32&0.68&-0.018\\
392&Brandon Ingram& \$31.651&45&6.60&-0.017\\
391&DeAndre Jordan& \$10.734&39&3.11&-0.017\\
390&Devin Booker& \$33.833&53&8.52&-0.017\\
389&Myles Turner& \$35.097&62&9.60&-0.017\\
388&Al Horford& \$26.500&63&6.92&-0.016\\
387&Kevin Durant& \$44.120&47&7.86&-0.016\\
386&Steven Adams& \$17.927&42&6.66&-0.016\\
385&Kira Lewis Jr.& \$4.004&25&0.91&-0.015\\
384&Gary Harris& \$13.000&48&3.17&-0.015\\
383&Robert Covington& \$12.308&48&3.33&-0.015\\
382&Doug McDermott& \$13.750&64&4.07&-0.015\\
381&Landry Shamet& \$9.500&40&2.83&-0.015\\
380&Chris Paul& \$28.400&59&7.52&-0.015\\
379&Jimmy Butler& \$37.653&64&10.40&-0.014\\
378&Jrue Holiday& \$34.320&67&9.82&-0.014\\
377&Nicolas Batum& \$19.700&78&5.70&-0.014\\
376&Jamal Murray& \$31.651&65&9.22&-0.014\\
375&Zach LaVine& \$37.097&77&11.04&-0.013\\
374&Giannis Antetokounmpo& \$42.492&63&13.62&-0.013\\
\hline
	\end{tabular}
}
\caption{
{\small
\textbf{Bottom 50 performers: Return on investment}.
The bottom 50 performers in terms of \eqref{eq:NBA_IRR} for the 2022-2023 NBA
regular season, based on a minimum of 25 games played.
For reference, we also include the player salary \citep{hoops_hype} in
millions, games played (GP), and PVGCP, as calculated with
\eqref{eq:pv_gcp}.
}
}
\label{tab:bottom50_ROI}
\end{table}

\begin{figure}[t!]
    \includegraphics[width=\textwidth]{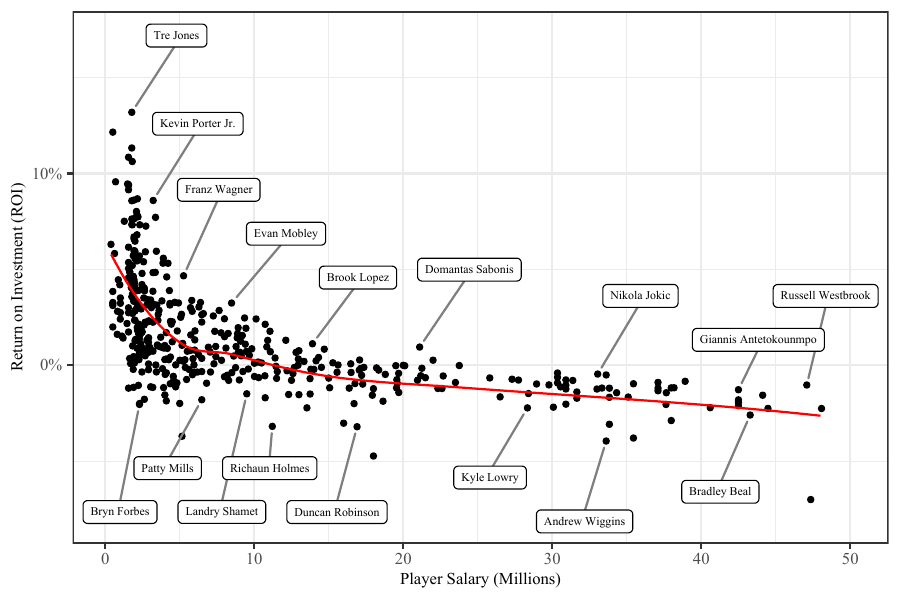}
\caption{
{\small
\textbf{Relative return on investment by salary}.
A scatter plot of contractual ROI by salary via \eqref{eq:NBA_IRR} for the
2022-2023 NBA regular season.  Because player salary is generally deterministic
by \citet{nba_cba}, this plot allows for relative comparisons within each
salary bracket.  For example, for players in the \$30M-\$35M range, Nikola
Joki\'{c} generated a higher relative regular season ROI than Andrew Wiggins.
The mean line was generated by the \texttt{loess} function in \citet{R_cite}.
The scatter plot shape is of interest, too, as it demonstrates maximum salary
players generally struggle to produce a break-even ROI based on regular
season performance only.
}
}
    \label{fig:roi_plot}
\end{figure}

As a final curiosity of the ROI framework we propose, it is of interest to
examine a scatter plot of ROI by salary.  In other words, player compensation in
the NBA is effectively formulaic and prescribed by the \citet{nba_cba}.
Thus, for players bracketed within certain salary ranges, it is useful to
identify which players generate the relative best contractual ROI.  We do
exactly this in Figure~\ref{fig:roi_plot}.  The best relative performers are
at the top of the resulting hockey stick shape, e.g., by increasing player
salary: Tre Jones, Kevin Porter Jr.,
Franz Wagner, Evan Mobley, Brook Lopez, Domantas Sabonis, Nikola Joki\'{c},
Giannis Antetokoumpo, and Russell Westbrook.  Conversely, the worse relative
performers are at the bottom of the hockey stick shape, e.g., Bryn Forbes,
Patty Mills, Landry Shamet, Richaun Holmes, Duncan Robinson, Kyle Lowry,
Andrew Wiggins, and Bradley Beal.  Furthermore, the overall shape of the
scatter plot in Figure~\ref{fig:roi_plot} is itself instructive.  Because it
is difficult for higher salary players to generate a break-even
ROI based only on regular season on court performance,
Figure~\ref{fig:roi_plot} implies that a considerable component of the
expectation for maximum salary players is play-off performance.  As a final
caveat, the results will likely change with a different single game metric,
different game values, or methods that go beyond just on court performance.
Further discussion may be found in Section~\ref{sec:disc}.

\section{Discussion}
\label{sec:disc}

The NBA is big business, and no small part is due to the over
\$4.4 billion in annual player compensation \citep{hoops_hype}.
Given the salary cap restrictions of the NBA \citep{nba_cba},
it is of paramount importance to team on
court success to appropriately compensate players for on court performance.
Despite this, there are no known studies that present a framework to
measure a player's contractual ROI.  This study is thus the first known
attempt.

\begin{figure}[t!]
    \includegraphics[width=\textwidth]{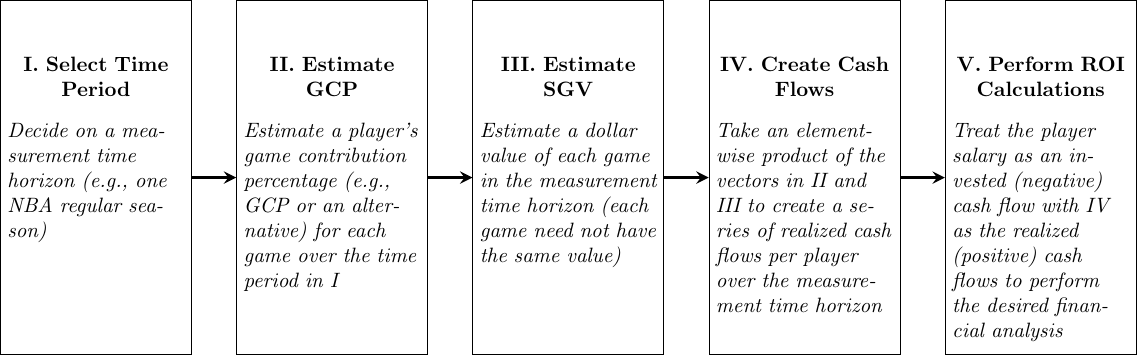}
\caption{
{\small
\textbf{NBA contractual ROI estimation framework summary}.
}
}
    \label{fig:roi_framework}
\end{figure}

Our approach unfolds in five parts.  We first decide on an investment time
horizon over which performance will be measured.  The next step is computing
a GCP metric.
(The GCP we propose in Section~\ref{subsec:GCP_method}, via
\eqref{eq:GCP_calc}, is itself a novel
contribution to the field of on court basketball player assessment metrics.
Because \eqref{eq:GCP_calc} is calculated per game,
cumulative metrics such as PVGCP, via \eqref{eq:pv_gcp},
allow analysts to assess the impact of missed games in a single calculation
(e.g., Figure~\ref{fig:gcp_comp}).
This has long been a known issue in the NBA \citep{wimbish_2023}. 
Additionally, PVGCP may offer a fresh perspective on player evaluation, given
the general consensus of the other popular player evaluation metrics reported
in Table~\ref{tab:top50_GCP}; i.e., PER, WS, BPM, VORP, and
RAPTOR unanimously ranking Joki\'{c} first, whereas PVGCP ranks Sabonis
first.)
After calculating GCP, the third step is to estimate the dollar value of
each NBA game in the measurement period (e.g., the SGV).  Fourth, the GCP
and SGV calculations are combined to convert a player's on court per game
performance into a series of realized cash flows.  From this, the fifth and
final step is to perform standard financial calculations by using the player's
salary as invested (i.e., negative) cash flows and the newly created income
(i.e., positive) cash flows from step four.  Our novel
framework is summarized in Figure~\ref{fig:roi_framework}.

The potential value of our proposed framework is illustrated in
Figure~\ref{fig:roi_plot}, which may be used by NBA player personnel decision
makers and NBA player agents alike in contract negotiations.  Additionally,
voters for NBA regular season awards may be interested in the results of
Table~\ref{tab:top50_GCP} or \ref{tab:top50_ROI}.  Indeed, the Kia NBA
Most Valuable Player award seems like a good candidate for the consideration
of PVGCP or ROI-type calculations.  In terms of forecasting,
it is not difficult to see how player on court projections may be used to
produce a distribution of GCP realizations, which may then be used to estimate
the dollar or trade value of draft picks or swaps or for potential trades more
generally.  Further, because player
contracts are highly regulated by the NBA CBA \citep{nba_cba}, the ROI
calculation methods herein may also be used for validation and fairness
purposes (e.g., Figure~\ref{fig:gcp_hist} and the break-even calculations of
Section~\ref{subsec:ROI_results} suggest the maximum salary restriction of
30-35\% of a team's salary cap appears reasonable).  GCP may also be used in
sports injury-related or performance-based studies.  For example,
\citet{page_2013} look at the effect of minutes played and usage on a player's
{\it production curve} over the course of their career.  Within the model,
the Game Score \citep{br_glossary} is used as a measure of production.  Our
GCP offers an alternative measure for a similar analysis.

In closing,
we again emphasize the main contribution of this study is a framework
to measure realized contractual ROI for NBA players.  As such, some simplifying
assumptions have been made, and it is possible our methodology may be
customized or enhanced. For example, the fields we select for the GCP
calculation in Table~\ref{tab:fields} are just one such proposal.  These may
be easily edited to meet the likely differing views of NBA analysts (NBA teams
may also possess more detailed player evaluation data than what is publicly
available, which is a further motivation for alternative field selections).
Further, in \eqref{eq:GCP_calc}, we use a simple,
uniform-like weighting system for the importance of each field in
Table~\ref{tab:fields}.  Alternative weighting schemes are also possible.  For
example, \citet{ozmen_2016} analyzes the marginal contribution of game
statistics across various levels of competitiveness in the Euroleague to win
probability.  Hence, a similar analysis could be utilized to vary the
weighting scheme of \eqref{eq:GCP_calc}, if desired.
Instead, additional precision may be
used to assign the weights within GCP, such as refining the quality of a
field-goal attempt \citep[e.g.,][]{shortridge_2014, daly_2019} or
accounting for peer (i.e., teammate) and non-peer (i.e., opponent) effects
\citep[e.g.,][]{horrace_2022}.  Even more, the GCP may be ignored altogether
and replaced with an alternative per game evaluation metric.  As long as the
percentage and per game properties hold, many alternatives to GCP are valid.

Beyond changes to the GCP metric, the ROI methods of
Section~\ref{subsec:ROI_method} may be enhanced or customized, too.  For
example, we assume the entire player salary is a time zero investment.
Instead, the actual payment dates of a player's salary may be used.  In the
same way, the actual game dates may be used instead of assuming 82 equally
spaced per game cash flows.  Further, we use a uniform weight for each game,
and the nature of \eqref{eq:NBA_IRR} implicitly weights early season games
more heavily than later season games.  As an alternative, it may be desirable
to assign different weights to each game based on its importance (i.e., the
proverbial ``big game'').  For example,
\citet{teramoto_2010} is an an example of how weighting schemes may differ
for playoff games versus regular season games in the NBA.  Or, to avoid
the implicit weighting of \eqref{eq:NBA_IRR}, it may be prudent to randomize
the order of the games and calculate a distribution of realized ROI
calculations.  Indeed, the SGV methodology of \eqref{eq:SGV} is quite
rudimentary and is thus ripe for additional study.
Furthermore, we consider only regular season games.  While this
is natural for regular season award considerations, there is the obvious
curiosity of how the calculations in Section~\ref{subsec:ROI_results} would
change with the inclusion of playoff games or even off court revenue, such as
jersey sales.
We close with a hopeful note in that the suggestions of this and the
previous paragraph may motivate additional study.

{\singlespacing
\footnotesize
\bibliographystyle{mcap}
\bibliography{refs}
}

\end{document}